\documentclass[aps,prb,twocolumn,groupedaddress,showpacs]{revtex4}
\usepackage{epsfig,amsfonts}

\def\k{{\bf k}}

\def\8{\infty}
\def\oh{\frac{1}{2}}

\def\d{\partial}

\def\undertext#1{\vtop{\hbox{#1}\kern 1pt \hrule}}

\def\tr{\hbox{tr}\,}

\def\pbyp#1#2{\frac{\partial#1}{\partial#2}}

\def\be{\begin{equation}}
\def\ee{\end{equation}}
\def\bea{\begin{eqnarray} & &}
\def\eea{\end{eqnarray}}

\def\rf#1{(\ref{#1})}

\def\cH{{\cal H}}
\def\rf#1{(\ref{#1})}

\def\rfs#1{Eq.~\rf{#1}}
\def\sign{{\rm sign}}

\begin{document}


\title{Single particle Green's functions and interacting topological insulators}


\author{V. Gurarie}
\affiliation{Department of Physics, University of Colorado,
Boulder CO 80309}


\date{\today}

\begin{abstract}
We study topological insulators characterized by the integer topological invariant $\mathbb Z$, in even and odd spacial dimensions. These are well understood in case when there are no interactions. 
We extend the earlier work on this subject  to construct their topological invariants in terms of their Green's functions. In this form, they can be used even if there are interactions. Specializing to one and two spacial dimensions, we further show that if two topologically distinct topological insulators border each other, the difference of their topological invariants is equal to the difference between the number of zero energy boundary excitations and the number of zeroes of the Green's function at the boundary. In the absence of interactions Green's functions have no zeroes thus there are always edge states at the boundary, as is well known. In the presence of interactions, in principle Green's functions could have zeroes. In that case, there could be no edge states at the boundary of two topological insulators with different topological invariants. This  may provide an alternative explanation to the recent results on one dimensional interacting topological insulators.

\end{abstract}

\maketitle

\section{Introduction}
The original proposals establishing existence of topological insulators with the $\mathbb Z_2$ topological invariant (in the presence of spin-orbit coupling with time-reversal invariance) \cite{Kane2005,Bernevig2006a} and their subsequent observations in experiment \cite{Konig2007,Hasan2008} contributed to a start of a tremendous amount of activity on this subject. 

Recently a classification table was worked out for all possible noninteracting topological insulators \cite{Ryu2010,Kitaev2009}. 10 symmetry classes of topological insulators were identified. In each spacial
dimensionality, representative systems of 5 classes are topologically trivial, while the remaining 5 classes can be topologically nontrivial.

The term ``topologically nontrivial" in this context implies the following. First of all, a noninteracting topological insulator, when coupled to an external electromagnetic field, is typically (but not always) expected to
have a response to this field given by a Chern-Simons term in even dimensional space or by a $\theta$-term in odd dimensional space \cite{Qi2008,Ryu2010}. Second, a topologically
nontrivial insulator has zero energy  excitations at its boundary (or at the boundary of two topologically distinct insulators). These are usually referred to as edge states. To derive all of these properties, typically one takes advantage of the fact that these systems are noninteracting fermions and thus are essentially exactly solvable. 

A natural question is then whether adding interactions to topological insulators preserves their observable properties, such as the existence and robustness of the edge states.

One particular approach to this problem was developed over 20 years ago in a series of papers by G. Volovik and collaborators, summarized in his book~\cite{VolovikBook1}. This approach  is based on the following simple idea. Noninteracting fermionic systems are described by a Hamiltonian which is equivalent to a matrix ${\cal H}_{ab}(\k)$, which is a function of momenta $\k$, as explained in Ref.~[\onlinecite{Ryu2010}]. This matrix satisfies certain conditions, which define the symmetry class of the topological insulator. For example, for insulators with the chiral (sublattice) symmetry, the condition reads
\be \label{eq:intrsym} {\cal H}(\k) = - \Sigma^\dagger {\cal H}(\k) \Sigma,
\ee where $\Sigma$ is a certain unitary matrix. Whether a particular system belongs to a non-trivial topological class can be investigated by studying the topological properties of the function ${\cal H}(\k)$ which defines a map from the space of the momenta $\k$ to the space of the matrices ${\cal H}$. 

Interacting systems are no longer described by  a matrix of this kind. However, one can define a Green's function (single particle retarded or advanced Green's function) for both interacting and noninteracting systems. For noninteracting systems, it takes especially simple form
\be G(\Omega, \k) = \left[ \Omega - {\cal H}(\k) \right]^{-1}.
\ee
The Green's function computed at zero frequency is simply equal to minus inverse of ${\cal H}$ and thus inherits all its symmetries, implying in case of  \rfs{eq:intrsym},
that
\be \label{eq:grrrr} G(0,\k) = - \Sigma^\dagger G(0,\k) \Sigma. 
\ee
At this point, we say that an interacting system belongs to a particular symmetry class if its Green's function $G(0, \k)$ satisfies the relations such as \rfs{eq:grrrr}.
Subsequently, we can study whether a particular interacting system is topological by studying the function $G(0,\k)$, instead of the earlier ${\cal H}(\k)$. 

All of this would stay in the realm of pure math, if it were not for a further observation in Ref.~[\onlinecite{VolovikBook1}], on the example of symmetry class D systems in 2 dimensions (class D is a shorthand for a superconductor without any symmetries, such as 2D $p+ip$ superconductor), that two topologically distinct systems (at least in two dimensional space) must have zero energy excitations at their boundary. This observation was made by examining the Green's function $G(\Omega,\k)$ only, without any reference to whether the system is interacting or noninteracting, although as we will see in this paper, an implicit assumption of the absence of interactions was indeed made.

The goal of this paper is to construct the framework which could in principle generalize these observations to arbitrary number of dimensions for all 10 symmetry classes. At the basis of this framework lies the construction of the topological invariants in terms of the Green's functions in arbitrary number of dimensions. One should note that simply rewriting the topological invariants in terms of $G^{-1}(0,\k)$ instead of $\cH(\k)$ is not sufficient. To be able to relate the topological invariant to the zero energy excitations one needs to have it written in space-time, as  topological characteristics of the map from $\Omega,\k$ space to $G$. 

Furthermore, in this work we are able to eliminate the implicit assumption of the absence of interactions. We do it by observing that two topologically distinct systems, in the presence of interactions, can have zeroes of the Green's functions at their boundary instead of the poles. The zero of the Green's function is defined as the zero of one of its eigenvalues, or equivalently of its determinant. If this indeed happens, two interacting topological insulators can have no edge states at the boundary between them and can be continuously deformed from one to another without ever closing the bulk gap. Of course, this picture does not explain when this scenario happens, rather it just says that it in principle can happen. 

In some cases one can give arguments that zero energy excitations cannot simply vanish as interaction is increased \cite{Ryu2010a}. These arguments, when they are at work, lead us to conclude that the poles cannot simply get replaced by the zeroes. On the other hand, it is also possible that while the edge states disappear as single particle excitations, they remain as collective excitations and continue to carry current. In this case, perhaps it is possible to reconcile the picture from Ref.~[\onlinecite{Ryu2010a}] and the one from this paper. 

At the same time, recent studies of interacting topological insulators in one dimensional space \cite{Fidkowski2010,Fidkowski2010a,Turner2010} show that in some cases the edge states do disappear as interactions are increased. To make contact with this work, in this paper we give the details of the Green's function construction in one dimensional space, for systems in symmetry class AIII or BDI \cite{Ryu2010}. 

The rest of the paper is organized as follows. In sec.~\ref{sec:2} we explain how to generalize the symmetry classes of topological insulators to the case when interactions are present. In Sec.~\ref{sec:3} we discuss the structure of poles and zeroes of the Green's functions. In Sec.~\ref{sec:4} we explain why the zeroes of the Green's function can change the topological invariants without any zero energy excitations. In Sec.~\ref{sec:vol} we discuss the application of this formalism to two dimensions for systems of the integer quantum Hall type, which mostly follows the derivation from Ref.~[\onlinecite{VolovikBook1}], but takes into account the possible zeroes of the Green's functions. This construction can be generalized to any symmetry class without chiral symmetry, which are classes A, AI, AII, D and C, in even number of spacial dimensions \cite{Ryu2010}.  Finally, Sec.~\ref{sec:6} we study chiral topological insulators in remaining classes AIII, BDI, CII, DIII, CI, in odd number of spacial dimensions. We especially concentrate on one dimensional AIII or BDI topological insulators. 
At the end of the paper in Sec.~\ref{sec:concl}, we discuss possible directions of further development of this formalism. 

\section{Green's functions of interacting topological insulators}

\label{sec:2}

We begin by  reviewing the symmetries of the noninteracting topological insulators. Those are characterized by matrix Hamiltonians $\cH_{ij}$. Here, and in the rest of this paper, the indices $i$, $j$ denote sites of a lattice, spin, and other degrees of freedom present in our problem. The total number of values $i$ takes is equal to the number of degrees of freedom in the problem, we will denote this $D_f$. Thus we take for the Hamiltonian of the problem
\be \hat H = \hat a^\dagger_i \cH_{ij} \hat a_j,
\ee
where $\hat a_i^\dagger$, $\hat a_i$ are fermionic creation and annihilation operators. 
In this formula, as well as in many others below, summation over the indices $i$, $j$ is implied (in some, but not all, cases we will write the summation symbol explicitly). We also note that sometimes in what follows $i$ will simply mean imaginary unit rather than an index; we hope the distinction will be clear from context. 

These may satisfy one of three constraints. The first is
time reversal invariance
\be \label{eq:tr} U_T^\dagger \cH^* U_T = \cH.
\ee
The second is
charge conjugation symmetry
\be \label{eq:cc} U_C^\dagger \cH^* U_C =- \cH.
\ee
Here $U_C$ and $U_T$ are unitary matrices, satisfying \be \label{eq:symp} U^*_{C} U_{C}=\pm 1, \ U^*_{T} U_{T}=\pm 1. \ \ee
In other words, each of these two symmetries, if it is present, can be implemented in two distinct ways, with either $+1$ in \rfs{eq:symp} or with $-1$.

In the presence of these two constraints, the Hamiltonian also automatically satisfies the constrain
\be \label{eq:sl}
\Sigma^\dagger \cH \Sigma = -\cH,
\ee where $\Sigma=U_C^* U_T$ is a unitary matrix. This is called chiral, or sublattice, symmetry. The sublattice symmetry can also be present in case when neither time reversal invariance or charge conjugation symmetry are present (although it cannot be present if only one of these two is missing). All together, taking into account \rfs{eq:symp}, one can see that there are exactly 10 separate symmetry classes of Hamiltonians, distinguished by the presence of absence of these constraints
\cite{Altland1997,Zirnbauer1996,Ryu2010,Kitaev2009}. 

Note that the Green's functions, defined by
\be G(\Omega) = \left[ \Omega- \cH \right]^{-1},
\ee
satisfy the following constraints in case when their parent Hamiltonians satisfy either of Eqs.~\rf{eq:tr}, \rf{eq:cc}, and \rf{eq:sl}. 
Time reversal invariance leads to
\be  \label{eq:gtr}
 G(\Omega)= U_T^\dagger G^T(\Omega) U_T ,
\ee
charge conjugation leads to
\be  \label{eq:gcc} G(\Omega)= - U_C^\dagger G^T(-\Omega) U_C,
\ee
(here the symbol $G^T$ denotes the transposed Green's function) and sublattice symmetry leads to
\be  \label{eq:gsl}  G(\Omega)=-\Sigma^\dagger G(-\Omega) \Sigma.
\ee

We would like to generalize the symmetries given by Eqs.~\rf{eq:tr}, \rf{eq:cc}, and \rf{eq:sl} to the case when interactions are present, in such a way that 
the interacting Green's function 
$G(\Omega)$ still satisfies either of the relations Eqs.~\rf{eq:gtr}, \rf{eq:gcc}, and \rf{eq:gsl}. In the rest of this section, we show that when properly generalized  to the presence of interactions, the symmetries of the Hamiltonian lead to exactly the same constraints on the Green's functions as in the absence of interactions \cite{Foster2006}. 

We first need to clarify what we mean by the interacting Hamiltonian and the interacting Green's function. The interacting Hamiltonian is constructed in terms of the creation and annihilation operators in the same way as the noninteracting one, except it is no longer quadratic in these operators. Interacting Green's function is then constructed in the following way. 

We introduce the retarded Green's function \cite{AGD} $G_{ij}(\Omega)$. It can be analytically continued onto the upper half plane of complex $\Omega$, where, on the purely imaginary axis, it becomes the imaginary time ordered Green's function $G_{ij}(i \omega)$,
where $\Omega=i\omega$ and $\omega$ will be taken to be real throughout this text. Since the system is supposed to be insulating and has a gap in the spectrum, $G_{ij}(\Omega)$ can also be continued onto the lower half plane through the gap where, just below the real axis, it becomes the advanced Green's function. For the real values of $\Omega$ where the density of states of the problem is not zero, the Green's function has cuts on the real axis through which it cannot be analytically continued (if it is, it will spill into another Riemannian sheet). In some cases we will be interested in topological insulators with boundary modes which will manifest themselves in excitations at zero and close to zero. In this case, to define the Green's function analytic everywhere we can think of a finite system so that the singularities of $G(\Omega)$ are confined to a number of poles. We will elaborate on the analytic properties of the Green's function in the next section. 

In addition to the Green's functions defined here, it is often customary to introduce time-ordered, anti-time ordered and Keldysh Green's functions \cite{Kamenev2004}. To distinguish them from the Green's functions defined in the previous paragraph, we will always specify which function we deal with whenever we work with one of those functions, retaining the name ``Green's function" for the analytic retarded-advanced-imaginary-time-ordered functions. 

Now let us generalize the symmetries of the Hamiltonian to the case when interactions are present. We define unitary operators $\hat U_T$, $\hat U_C$ and $\hat \Sigma$, which act on creation and annihilation operators $\hat a^\dagger_i$ and $\hat a_i$
(where $i$ labels the sites of a lattice as well as other degrees of freedom if present) in the following way
\be  \label{eq:trio} 
\hat U_T^\dagger \hat a_i \hat U_T = \sum_j {U_T}_{ij} \hat a_j, \ U_T^\dagger \hat a_i^\dagger \hat U_T = \sum_j  \hat a^\dagger_j {U_T}^\dagger_{ji},
\ee
\be \label{eq:pho} \hat U_C^\dagger \hat a_i \hat U_C = \sum_j U^\dagger_{C ij} \hat a^\dagger_j, \ \hat U_C^\dagger \hat a_i^\dagger \hat U_C = \sum_j  \hat a_j U_{C ji},
\ee
and
\be \label{eq:chio} \hat \Sigma^\dagger \hat a_i \hat \Sigma = \sum_j  \hat a^\dagger_j \Sigma^\dagger_{ji},\
\hat \Sigma^\dagger \hat a_i^\dagger \hat \Sigma = \sum_j  \hat  \Sigma_{ij} \hat a_j.
\ee
We now declare that a many body Hamiltonian $\hat H$ is time reversal invariant if
\be\label{eq:tri} \hat U_T^\dagger \hat H^* \hat U_T = \hat H, 
\ee
is particle-hole symmetric if
\be \label{eq:phtrans} \hat U_C^\dagger \hat H \hat U_C=\hat H,
\ee
and is chiral if
\be \label{eq:chi}  \hat \Sigma^\dagger \hat H^* \hat \Sigma =\hat H.
\ee
Here $\hat H^*$ is a complex conjugate many body Hamiltonian. The creation and annihilation operators are not affected by the operation of complex conjugation. 

Now it is straightforward to check that if the Hamiltonian is noninteracting, that is if 
\be \hat H = \sum_{ij} \hat a^\dagger_i {\cH}_{ij} \hat a_j,
\ee
then Eqs.~\rf{eq:tri} and \rf{eq:trio} are equivalent to \rfs{eq:tr}, Eqs.~\rf{eq:phtrans} and \rf{eq:pho} are equivalent to \rfs{eq:cc}, while Eqs.~\rf{eq:chi} and \rf{eq:chio} are equivalent to \rfs{eq:sl}. For example,
\begin{eqnarray} \hat U_T^\dagger \hat H^* \hat U_T &=& \hat U_T^\dagger \hat a_i^\dagger \cH^*_{ij} \hat a_j \hat U_T = 
\hat  a^\dagger_k U^\dagger_{T ki} \cH^*_{ij} U_{T jl} \hat a_l= \cr
\hat a^\dagger_i \cH_{ij} \hat a_j &=& \hat H.
\end{eqnarray}
Similarly,
\begin{eqnarray} \hat U_C^\dagger \hat H \hat U_C& =& \hat U_C^\dagger \hat a_i^\dagger \cH_{ij} \hat a_j \hat U_C = \hat a_k U_{Cki} \cH_{ij} U^\dagger_{Cjk} \hat a^\dagger_k=\cr
-\hat a_j \cH^*_{ji} \hat a_k &=& \hat a^\dagger_i \cH_{ij} \hat a_j
\end{eqnarray}
 (this assumes, without loss of generality, that $\tr \cH =0$), and finally
\begin{eqnarray} \hat \Sigma^\dagger \hat H^* \hat \Sigma &=& \hat \Sigma^\dagger \hat a_i^\dagger {\cal H}^T_{ij} \hat a_j \hat \Sigma = \Sigma_{ik} \hat a_k {\cal H}^T_{ij} \hat a^\dagger_l \Sigma^\dagger_{lj} =\cr -
\hat a^\dagger_l \Sigma^\dagger_{lj} {\cal H}_{ji} \Sigma_{ik} \hat a_k &=& \hat a^\dagger_i {\cal H}_{ij} \hat a_j = \hat H.
\end{eqnarray}

The generalizations given here are not the only way to generalize these symmetries to the case when interactions are present. One alternative way to generalize the chiral symmetry is given in appendix~\ref{sec:appendixA}. However, they seem to be most natural from the point of view of realistic applications.

Next, one can check that even in the presence of interactions Eqs.~\rf{eq:gtr}, \rf{eq:gcc} and \rf{eq:gsl} still hold. This is a straightforward yet  tedious calculation which can be skipped on the first reading. 

We work with the time ordered and anti-time ordered Green's functions
\begin{eqnarray} G^t_{ij}(t) &=&-i \left< 0 \right|T  \hat a_i(t) \, a^\dagger_j(0) \left| 0 \right>, \cr
G^{\tilde t}_{ij}(t) &=&-i \left< 0 \right|\tilde T  \hat a_i(t) \, a^\dagger_j(0) \left| 0 \right>.
\end{eqnarray}
Here the symbols $T$ and $\tilde T$ denote time and anti-time ordering of subsequent time dependent operators. We denote the corresponding Green's functions $G^t$ and $G^{\tilde t}$ (to avoid confusion with the notation $G^T$ which implies transposed Green's function $G^T_{ij} = G_{ji}$). Given these time ordered Green's functions, we can construct regarded and advanced Green's functions according to
\begin{eqnarray} G^R_{ij}(t) &=&  \theta(t) \left( G^t_{ij}(t) -G^{\tilde t}_{ij}(t) \right), \cr  G^A_{ij}(t) &=& \theta(-t) \left( G^t_{ij}(t) -G^{\tilde t}_{ij}(t) \right).
\end{eqnarray}
Fourier transform of those gives us either $G^R(\Omega)$ with complex  $\Omega$ in the upper half plane or $G^A(\Omega)$ with complex $\Omega$ in the lower half plane, which are known to be analytic continuation of each other (at least in the case of the gapful spectrum), and define the same analytic function $G(\Omega)$ in the entire complex plane. To prove that thus defined $G(\Omega)$ satisfies 
Eqs.~\rf{eq:gtr}, \rf{eq:gcc} or \rf{eq:gsl}, depending on the symmetries present, it is sufficient to prove that the time ordered and anti-time ordered Green's function satisfy one of the three relations,
\be  \label{eq:gtr1}
 G^t(t)= U_T^\dagger {G^t}^T(t) U_T ,
\ee
\be  \label{eq:gcc1} G^t(t)= - U_C^\dagger {G^t}^T(-t) U_C,
\ee and
\be  \label{eq:gsl1}  G^t(t)=-\Sigma^\dagger G^t(-t) \Sigma
\ee
(same for the anti-time ordered Green's function). 

Start with the time reversal invariance. The presence of the symmetry \rfs{eq:tri} implies that for every many-body state $\left| n \right>$, there exists a state
$\hat U_T^\dagger\left| n^* \right>$ which has exactly the same energy as the state $\left| n \right>$ (here $\left| n^* \right>$ denotes the complex conjugate state). Indeed, assuming that $\hat H \left| n \right> = E_n \left| n \right>$ gives
\be \hat H \hat U_T^\dagger \left| n^* \right> = \hat U_T^\dagger \hat U_T \hat H \hat U_T^\dagger \left| n^* \right> = \hat U_T^\dagger \hat H^* \left| n^* \right> =
E_n \hat U_T^\dagger \left| n^*\right>. 
\ee
Without loss of generality we assume that the ground state is time reversal invariant, that is
\be \label{eq:statetr}  \hat U_T^\dagger \left| 0^* \right> = \left| 0 \right>, \ \left| 0^* \right> = \hat U_T \left| 0 \right>.
\ee
If it were not time reversal invariant, we could still take expressions such as $\left| 0 \right> \pm \hat U_T^\dagger \left| 0^* \right>$ as a ground state, which are invariant. 

Now assume that $t>0$. Then the following sequence of transformations can be carried out
\begin{eqnarray} \label{eq:mantr} 
&& iG^t_{ij}(t) = \left< 0 \right| e^{i \hat H t} \hat a_i e^{-i \hat H t} \hat a^\dagger_j \left| 0 \right> = \cr && \left< 0 \right| \hat U_T^\dagger e^{i \hat H^* t} \hat U_T \hat a_i U^\dagger_T e^{-i \hat H^* t} \hat  U_T \hat a^\dagger_j \hat U_T^\dagger \hat U_T \left| 0 \right>  =\cr &&
\left< 0^* \right| e^{i \hat H^* t} \hat U_T \hat a_i U^\dagger_T e^{-i \hat H^* t} \hat  U_T \hat a^\dagger_j \hat U_T^\dagger  \left| 0^* \right> =\cr
&& U^\dagger_{Tik}  \left< 0^* \right| e^{i \hat H^* t} \hat a_k e^{-i \hat H^* t} \hat a^\dagger_l  \left| 0^* \right> U_{lj} =
\cr && U^\dagger_{Tik}  \left< 0  \right| \hat a_le^{-i \hat H t} \hat a_k^\dagger e^{i \hat H t}   \left| 0 \right> U_{lj}  = \cr
&& U^\dagger_{Tik} \, {iG^t}^T_{kl}(t) U_{Tlj}. 
\end{eqnarray}
To go from the third line to the fourth line of  \rfs{eq:mantr} we took advantage of \rfs{eq:statetr}. To go from the third to the fourth line, we used that for general states $\left|n\right>$ and $\left| m \right>$ and for generic operator $\hat A$, the following is obviously true
\be \label{eq:transp} \left< n \right| \hat A \left| m \right> = \left< m^* \right| \hat A^T \left| n^* \right>.
\ee
We could also prove a similar relation for $t<0$, but this is not really necessary since we can just use the retarded Green's functions to construct $G(\Omega)$. At
the same time, we can easily prove the same for the anti-time ordered function $G^{\tilde t}$. 
 Thus we have proved \rfs{eq:gtr1} and, by extension, \rfs{eq:gtr}.
 
 We now move on to charge conjugation symmetry. Just as before, each excited state is doubly degenerate, and the ground state satisfies
 \be \hat U_C \left| 0 \right> = \left| 0 \right>.
 \ee
 We now do the following sequence of transformations for $t>0$
 \begin{eqnarray} \label{eq:mancc}
 && iG^t_{ij}(t) = \left< 0 \right| e^{i \hat H t} \hat a_i e^{-i \hat H t} \hat a^\dagger_j \left| 0 \right> = \cr && 
  \left< 0 \right| \hat U_C^\dagger e^{i \hat H t}  \hat U_C \hat a_i \hat U_C^\dagger e^{-i \hat H t} \hat U_C \hat a^\dagger_j \hat U_C \hat U_C^\dagger \left| 0 \right> = \cr
  && U_{Cjl}   \left< 0 \right| e^{i \hat H t}   \hat a_k^\dagger  e^{-i \hat H t}  \hat a_l \left| 0 \right>  U^\dagger_{Cki} = \cr
  && - U_{Cjl}   \left< 0 \right|T   \hat a_l e^{i \hat H t}   \hat a_k^\dagger  e^{-i \hat H t} \left| 0 \right>  U^\dagger_{Cki} = \cr
&&  - U_{Cjl}\,  iG^t_{lk}(-t) U^\dagger_{Cki}
 \end{eqnarray}
 From this \rfs{eq:gcc1} easily follows. In addition, same relations hold for negative time $t<0$ as well as for the anti-time ordered Green's functions $G^{\tilde t}$.
Thus we have proved 
\rfs{eq:gcc}. 

Finally, for the chiral (sublattice) symmetry, we repeat the same process. We take the ground state to be invariant under the chiral transformation, which
leads to
\be \label{eq:grchi} \hat \Sigma^\dagger \left| 0^* \right> = \left| 0 \right>, \ \left| 0^* \right> = \hat \Sigma \left| 0 \right>.
\ee

 We find, for $t>0$, 
\begin{eqnarray}
 && iG^t_{ij}(t) = \left< 0 \right| e^{i \hat H t} \hat a_i e^{-i \hat H t} \hat a^\dagger_j \left| 0 \right> = \cr 
  && \left< 0 \right| \Sigma^\dagger e^{i \hat H^* t} \hat \Sigma \hat a_i \hat \Sigma^\dagger e^{-i \hat H^* t} \hat \Sigma \hat a^\dagger_j \hat \Sigma^\dagger \hat \Sigma \left| 0 \right> = \cr && \Sigma^\dagger_{ik} \left< 0^* \right| e^{i \hat H^* t} \hat a^\dagger_k  e^{-i \hat H^* t} \hat a_l   \left| 0^* \right>  \Sigma_{lj}=\cr
 && \Sigma^\dagger_{ik} \left< 0 \right| \hat a_l^\dagger e^{-i \hat H t} \hat a_k  e^{i \hat H t}    \left| 0 \right>  \Sigma_{lj}= \cr 
&&  - \Sigma^\dagger_{ik} \left< 0 \right| T e^{-i \hat H t}  \hat a_k  e^{i \hat H t} \hat a^\dagger_l   \left| 0 \right>  \Sigma_{lj} =
\cr
&& - \Sigma^\dagger_{ik} \, {iG^t}_{kl}(-t) \Sigma_{lj}.
\end{eqnarray}
\rfs{eq:gsl1} follows immediately, and, together with the extension of this relation for $t<0$ and for $G^{\tilde t}$, \rfs{eq:gsl} follows as well. 

Thus we have proven that Eqs.~\rf{eq:gtr}, \rf{eq:gcc}, and \rf{eq:gsl} hold even in the presence of interactions. 

Finally, note that regardless whether or not the system is interacting, its Green's function satisfies
\be G_{ij}(\Omega)=G^\dagger_{ij}(\Omega^*).
\ee
Here $G^\dagger$ is the Hermitian conjugate of the Green's function understood as a matrix. 

\section{Analytic properties of the Green's functions}
\label{sec:3}

In this section, we discuss the analytic properties of the Green's functions, important for subsequent applications. The main result of this section is that the determinant of the Green's function can always be written in the following form
\be \label{eq:det} \det G = \frac{\prod_{s=1}^{D_h -D_f} \left(\Omega-r_s \right)}{\prod_{n=1}^{D_h} \left( \Omega- \epsilon_n \right)}.
\ee
Here $r_s$ are  real numbers representing positions of the zeroes of the determinant, $\epsilon_n$ are the excitation energies which are also obviously real and which represent the poles of the determinant, $D_h$ is the combined dimension of the Hilbert spaces of the system with one extra particle than the ground state and one less particle than the ground state, and $D_f$ is the number of degrees of freedom, or number  of distinct creation or annihilation operators as discussed earlier. Generally \be D_f \le D_h, \ee where the equality is achieved, in particular, if the system is not interacting. This can be seen if one notes that in the presence of interactions $D_h$ consists of all the states obtained from the ground state by adding or removing a particle (which can be done in $N_f$ distinct ways) plus states which differ from those by  particle-hole excitations. 

If the system is not interacting, $D_h=D_f$ and the determinant of its Green's function is particularly easy to calculate starting from $G= \left[ \Omega - \cH \right]^{-1}$, to give
\be \label{eq:detnonint} \det G = \prod_{n=1}^{D_f} \frac{1}{\Omega- \epsilon_n},
\ee which is consistent with \rfs{eq:det}. 

In the presence of interactions, the determinant of the Green's functions, in addition to its poles, has zeroes as follows from \rfs{eq:det}. Unlike the poles which have clear physical meaning as excitation energies, the zeroes are more subtle and, as we will see below, their positions reflect the matrix elements of the creation and annihilation operators. However, it should be clear that as interactions of the system are switched off, the zeroes move to coincide with the positions of some of the poles, resulting in disappearance of one zero and one pole at a time from the determinant, until in the absence of interactions it takes the form \rfs{eq:detnonint}. Conversely, if the interactions are turned on, the zeroes and poles appear in pairs. 

In the rest of this section we present the proof of \rfs{eq:det}. This can be skipped on the first reading. 
We begin by writing down the  spectral decomposition (sometimes referred to as Lehmann decomposition) of the Green's function
\be \label{eq:grleh} G_{ij}(\Omega) = \sum_n \frac{\left< 0 \right| \hat a_i \left| n \right> \left< n \right| \hat a^\dagger_j \left| 0 \right>}{\Omega- \omega^+_n} +
\sum_m  \frac{\left< 0 \right| \hat a^\dagger_j \left| m \right> \left< m \right| \hat a_i \left| 0 \right>}{\Omega- \omega^-_m}.
\ee
Here $\left| n \right>$ are the states which have one more particle compared to the ground state, and
\be \label{eq:omegaplus} \omega^+_n = E_n(N+1)-E_0(N)\ge 0,
\ee where $E_0(N)$ is the ground state of the system of $N$ particles (we take the chemical potential as included in the Hamiltonian, thus $dE_0(N)/dN=0$), and $E_n(N+1)$ is the excited state of a system with one more particle. Similarly, $\left| m \right>$ have one less particle than the ground state, and
\be \label{eq:omegaminus} \omega^-_m = E_0(N)-E_m(N-1)\le 0.
\ee 
It is convenient for the moment to work with combined notations, where  all states are denoted by $\left| n \right>$ regardless or not whether they have one more particle or one less particle than the ground state and all the energy is denoted by $\epsilon_n$ which can be either positive or negative and coincide with either $\omega_n^+$ or $\omega^-_m$. Then we can write
\be \label{eq:lehmm} G_{ij}(\Omega) = \sum_n \frac{U^\dagger_{in} U_{nj}}{\Omega - \epsilon_n},
\ee
where $U_{nj} = \left< n \right| \hat a^\dagger_j \left|0 \right>$ if $\left|n \right>$ has one more particle compared to the ground state and $U_{nj} = \left< 0 \right|
\hat a^\dagger_j \left| n \right>$ if $\left|n \right>$ has one less particle. 
Note that
\be \sum_n U^\dagger_{in} U_{nj} = \delta_{ij}.
\ee
In the matrix $U_{nj}$, $n$ goes over exactly $D_h$ values while $j$ goes over $D_f$ values, where, as we recall, $D_f\le D_h$. 

Now we are in the position  to study the determinant of the Green's function $\det G$, to derive \rfs{eq:det}.


If the system is non-interacting, then it has as many excited states as the number of its degrees of freedom or $D_f=D_h$. Then $U_{nj}$ is a unitary matrix which can be diagonalized. It follows that the eigenvalues of the Green's functions are
\be \lambda_\mu = \frac{1}{\Omega - \epsilon_\mu },
\ee
so that we immediately reproduce \rfs{eq:detnonint}.

If, however, the system is interacting, then $D_h$ can be larger than $D_f$ and $U_{nj}$ is a rectangular matrix. In that case, we write down the formal
expression for the determinant of the Green's function, starting from the definition \rfs{eq:lehmm}
\be
\det G =\sum_\sigma \sum_{n_1, n_2, \dots} \prod_i \frac{U^\dagger_{i n_i} U_{n_i\sigma(i)}}{ \Omega-\epsilon_{n_i} } (-1)^{P(\sigma)}
\ee
Here as usual  $\sigma(i)$ is a permutation of the numbers $1, 2, \dots, N_f$, and $P$ denotes its parity. 

It is now easy to see that if any two $n$ coincide, for example if $n_1=n_2$, then the summation over permutations gives zero. This means that the determinant of the Green's function
has simple poles at each point $\epsilon_n$ (and does not have any higher order poles). This justifies the denominator of \rfs{eq:det}. 

To explain the origin of the numerator, we first note that the determinant of the Green's function is clearly a rational function of $\Omega$, thus the numerator must be a polynomial. Moreover, it is well known \cite{AGD} that at very large $\Omega$ the Green's function asymptotically approaches a very simple form $G_{ij} \sim \delta_{ij}/\Omega$. This means that its determinant must go as $1/\Omega^{D_f}$. This implies that the polynomial in the numerator should be of the order $D_h-D_f$, with the highest order term $\Omega^{D_h-D_f}$ having the coefficient equal to unity. 

Thus the numerator of the determinant must take precisely the form shown in \rfs{eq:det}. The only thing which we have not yet proven is that all $r_s$ are real.

To prove that, let us prove that the eigenvalues of the Green's function cannot have zeroes (cannot vanish) when $\Omega$ is in the upper half plane or in the lower half plane.  In this proof we follow the arguments given in Refs.~[\onlinecite{AGD,LL5}]. Let us examine  the eigenvalue equation
\be  \label{eq:eigenvaluegr}
 \sum_{jn} \frac{U^\dagger_{in} U_{nj}}{\Omega - \epsilon_n} \psi_j^{(\mu)}(\Omega) = \lambda_\mu(\Omega) \psi^{(\mu)}_i(\Omega).
 \ee
 Then for $\Omega=E$ where $E$ is a real number which makes the Green's function hermitian
 \be \label{eq:eigenreps} \lambda_\mu (E) = \sum_n \frac{\left| \sum_j U_{nj} \psi_j^{(\mu)}(E) \right|^2 }{E - \epsilon_n} =
 \sum_n \frac{Z_n^{(\mu)}}{E - \epsilon_n},
 \ee
where the notation
 \be \label{eq:positivity} Z_n^{(\mu)}(E) = \left| \sum_j U_{nj} \psi_j^{(\mu)}(E) \right|^2 \ge 0
 \ee
 was introduced.
In addition, if $\Omega$ is taken to infinity, $\lambda_\mu(\Omega)\sim 1/\Omega$ as we have already discussed.

\begin{figure}[ht]
\includegraphics[height=3 in]{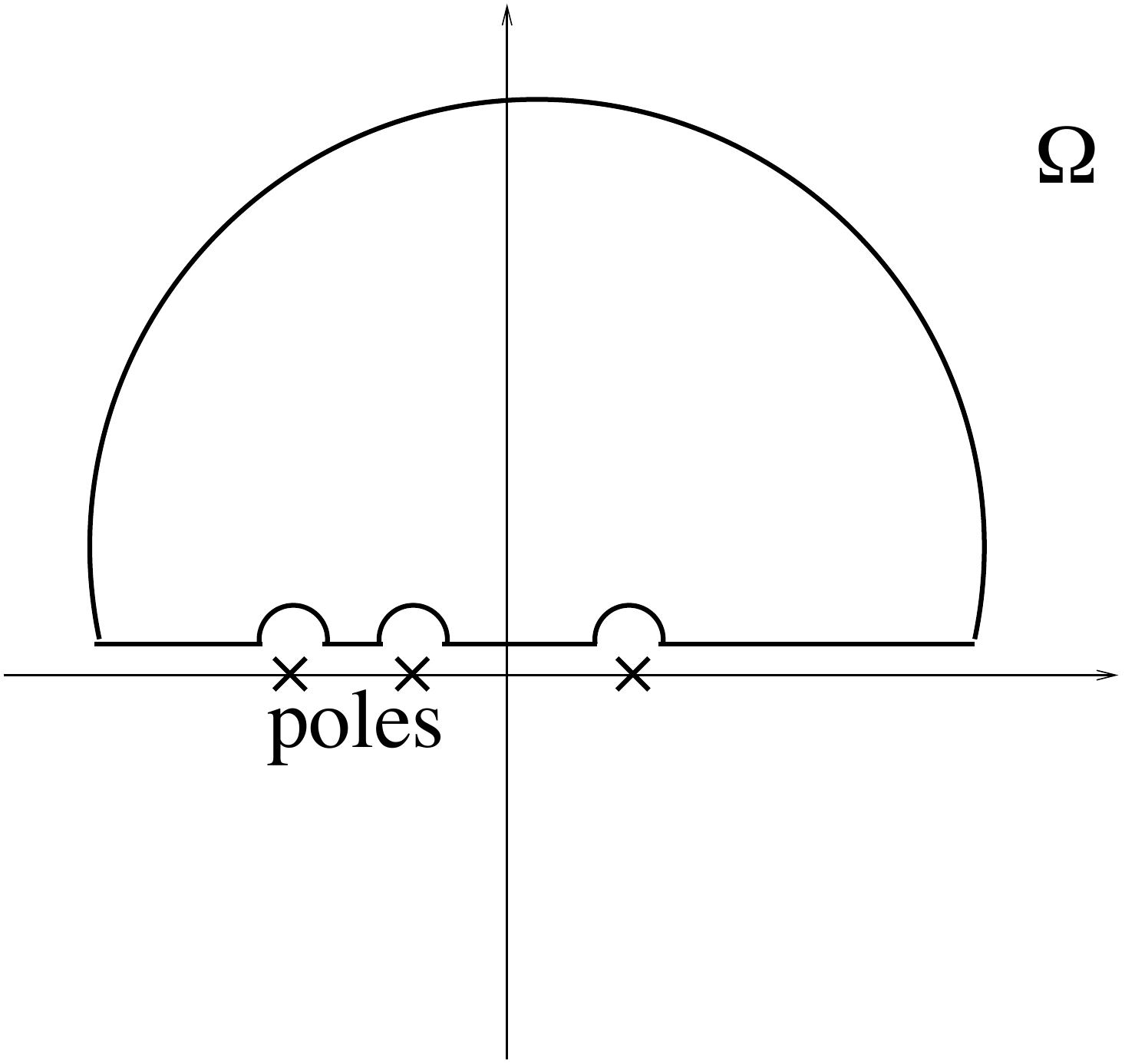}
\caption{The contour $\Omega$ is taken along in order to prove that the eigenvalues of the Green's functions cannot vanish in the upper half plane.}
\label{fig:contour}
\end{figure}

Now let us take $\Omega$ around the contour shown on Fig.~\ref{fig:contour}. The horizontal part of the contour goes infinitesimally above the real axis. On this contour, it follows from Eqs.~\rf{eq:eigenreps} and \rf{eq:positivity} that the imaginary part of an eigenvalue $\lambda_\mu(\Omega)$ is always zero or negative (negative if $\Omega$ is close to one of $\epsilon_n$). At the same time, when $\Omega$ goes over a large semicircle, $\lambda_\mu(\Omega) = 1/\Omega$ and its imaginary part is also negative. Thus ${\rm Im} \, \lambda_\mu(\Omega) \le 0$ for all $\Omega$ belonging to the contour shown on Fig.~\ref{fig:contour}. 

However, there exist a well known theorem in complex analysis which states that if an analytic function $\lambda(\Omega)$ has poles or zeroes in a certain domain, then if one plots $\lambda(\Omega)$ as $\Omega$ encircles this domain counterclockwise, the number of times $\lambda(\Omega)$ encircles zero  in the counterclockwise direction is equal to the number of zeroes minus the number of poles of this analytic function inside this domain. We see that as $\Omega$ follows 
the contour shown on Fig.~\ref{fig:contour}, $\lambda_\mu(\Omega)$ does not encircle zero at all. Since $\lambda_\mu(\Omega)$ cannot have any poles in the upper half plane, it follows that it cannot have any zeroes in the upper half plane.

Similarly, $\lambda_\mu(\Omega)$ cannot have any zeroes also in the lower half plane. This means they can have zeroes only on the real axis. 

If the eigenvalues of a matrix have zeroes on the real axis only, this implies that the determinant of the matrix can have zeroes on the real axis only as well.
Then the fact that all $r_s$ in \rfs{eq:det} are real follows. 

This concludes our proof of \rfs{eq:det}.

 \section{Mode counting via the Green's functions}
 \label{sec:4}
 
 At the heart of the approach to the topological insulators developed by Volovik \cite{VolovikBook1} lies the following topological invariant
 \be \label{eq:simplestinvariant}
 N_1 = {\rm Tr} \, \int_{-\infty}^\infty \frac{d\omega}{\pi i} \, K \d_\omega G.
 \ee  
 Here, first of all,  $\omega$ is a real variable such that $\Omega=i \omega$.  Second, $K_{ij}$ is a matrix inverse to $G_{ij}$
 \be \label{eq:trueinverse} \sum_j K_{ij} G_{jl} =\delta_{il}.
 \ee
 The notation $G^{-1}$, which would seem to be natural to use instead of $K$ in this instance, is going to be reserved for a different quantity which we will define later. 
 
The existence of this topological invariant reflects the fact that $\pi_1(GL(D_f,{\mathbb C}))={\mathbb Z}$. It is straightforward to see that if the system is perturbed and its Green's function changes by small amount as $G \rightarrow G + \delta G$, this quantity remains unchanged. 
 
 To see the meaning of this quantity, we express it in terms of the eigenvalues of the Green's function
 \be N_1 = \sum_\mu  \int_{-\infty}^{\infty} \frac{d\omega}{\pi i} \pbyp{\ln \left(\lambda_\mu \right)}{\omega} = \int_{-\infty}^\infty \frac{d \omega}{\pi i}
 \pbyp{\ln \det G}{\omega}.
 \ee
 It is easy to see by direct substitution of \rfs{eq:det} that the invariant is equal to
 \be \label{eq:posneg} N_1 = \sum_n {\rm \sign}\, \epsilon_n - \sum_s {\rm sign}\, r_s.
 \ee
that is  the difference between the total number of positive and negative poles, minus the difference between the total number of positive and negative zeroes of the determinant of the Green's function.

For noninteracting systems, $N_1$ is simply the difference between the number of positive and negative eigenvalues of the system \cite{VolovikBook1}, as noninteracting systems have no zeroes. For interacting systems, $N_1$ turns out to be more complicated. Notice, however, that $N_1$ can change only if a pole crosses zero (it increases by one if the pole moves from negative to positive values), or if a zero crosses zero (it increases by one  if a zero moves from positive to negative values). 

\begin{figure}[ht]
\includegraphics[height=.9 in]{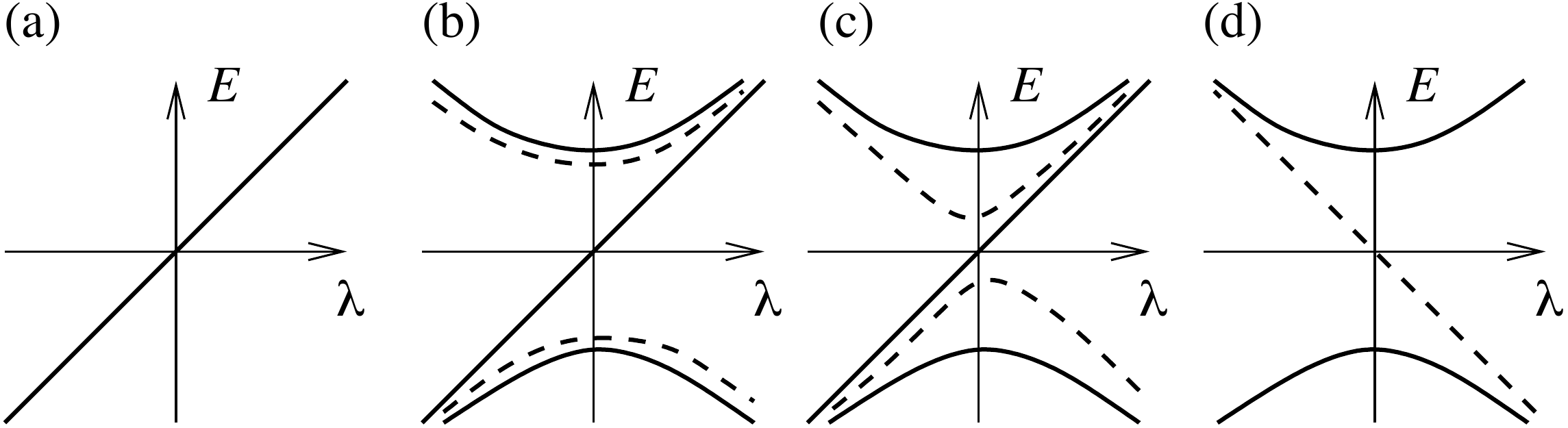}
\caption{A possible scenario outlining the change in the position of the poles (solid lines) and zeroes (dashed lines) as a function of some parameter $\lambda$, as the interaction strength is varied. (a) an energy level crossing zero as a function of $\lambda$ for a noninteracing system. (b) Two zeroes and and two nearby poles appear in the determinant of the Green's function \rfs{eq:det} as the interactions are increased (c) zeroes move closer to the positions of the  original poles of the noninteracting problem (d) zeroes ``annihilate" the poles resulting in  a system with exactly the same change of the topological invariant, but without any energy levels crossing zero energy.}
\label{fig:polemoves}
\end{figure}

This immediately leads to the following conclusion. Suppose we have an insulating system (it has a gap in the spectrum). If the system is noninteracting, the only way it can increase the number of positive energy levels is by moving an energy level from negative to positive values (and thus closing the gap somewhere in the middle of the process where this energy level crosses zero). If the system is interacting, it can change the number of positive energy levels by moving a zero from positive to negative values (where it eventually may ``fuse" with a pole and remove it from the spectrum). This process happens without ever closing the gap. A possible scenario outlining the motion of zeroes and poles is shown on Fig.~\ref{fig:polemoves}. 

This observation lies at the core of the possible transitions between two topologically distinct insulating states without closing the gap, in the presence of interactions. We will further elaborate on it later in this paper. 

\section{Topological insulators in two dimensional space}
\label{sec:vol}

In two spacial dimensions there are three types of topological insulators characterized by the topological invariant ${\mathbb Z}$.   These are class A (an example is integer quantum Hall effect), class D (an example is a $p+ip$ superconductor) and class C (an example is a 2D superconductor which is not time reversal invariant but with spin rotation invariance) \cite{Altland1997,Ryu2010}. More generally, representative systems of a total of 5 symmetry classes can possess a $\mathbb Z$ invariant in even spacial dimensions. 
 
The topological invariant which characterizes these systems in the context of integer quantum Hall effect in the absence of interactions is referred to as TKNN invariant \cite{Thouless1982}. It is usually written as a flux of the Berry connection through the Brillouin zone. However it is also possible to write it in terms of the Green's functions, as was first observed in Ref.~[\onlinecite{Niu1985}]. The advantage of the expression in terms of Green's functions is that it is then straightforward to generalize it to the case when interactions are present. 

To write down this invariant, we first introduce the Green's function in the momentum space, $G_{ab}(\Omega, \k)$ (this assumes that our system
is translationally invariant). Here the indices $a$, $b$ label the remaining degrees of freedom once momentum is introduced (these can be the basis of the lattice, spin, and other degrees of freedom). We also introduce the notations $k_0 = \omega$, where $\Omega = i\omega$ and $\omega$ is real, in addition to the momentum vector components $k_1$ and $k_2$. 

 Now the TKNN invariant takes the form \cite{Niu1985}
\begin{eqnarray} \label{eq:TKNN} && N_3 = \\ &&
\frac{\epsilon_{\alpha \beta \gamma}}6  \tr \int_{-\infty}^\infty d\omega  \int \frac{d^2 k}{(2\pi)^2} \, G^{-1} \d_{k_\alpha} G G^{-1} \d_{k_\beta} G G^{-1} \d_{k_\gamma} G, \nonumber
\end{eqnarray}
where summation over $\alpha$, $\beta$, $\gamma$ taking values $0$, $1$, and $2$ is implied, and $\tr$ denotes the trace over the matrix indices of $G$. $G^{-1}$ is a matrix inverse of $G$, defined as in
\be  \sum_b G^{-1}_{ab}(\Omega,\k) \, G_{bc}(\Omega, \k) = \delta_{ac}.
\ee
Following Ref.~[\onlinecite{VolovikBook1}], we denote this invariant by $N_3$, to emphasize its similarity with \rfs{eq:simplestinvariant}. 
The value of this invariant is an integer number which, in case of integer quantum Hall effect, is proportional to the Hall conductivity $\sigma_{xy}$. 

However, we can now imagine interactions to our system. The presence of interactions does not change the fact that $N_3$ is a topological invariant. Indeed, its existence reflects the homotopy group $\pi_3(GL(M,{\mathbb C})) = {\mathbb Z}$ for a matrix of a size $M$ by $M$ and is totally unrelated to whether the Green's function describes interacting or noninteracting system. 

However, a question still remains whether the existence of this topological invariant in a system with interactions would have any consequences for its physical properties. In Ref.~[\onlinecite{VolovikBook1}] it was shown that in the absence of interactions, if there are two domains separated by a straight one dimensional domain wall where $N_3$ takes on different values on either side of the domain wall, this implies that there are zero energy excitations on the domain wall, that is, the edge states. 

Let us see what remains of this phenomenon  in the presence of interactions. We will see that if two domains with two different values of the invariant share a domain wall which separates them, then even in the presence of interactions one can have zero energy excitations on the domain wall. However, the new effect which appears in the presence of interactions is that instead of a zero energy excitation, the Green's function can also develop a zero at zero energy on the domain wall. This would still lead to the change in topological invariant from the first to the second domain. Thus in the presence of interactions, sometimes it is possible to have to adjacent distinct topological insulators without any edge states in between. 

To derive this result, we closely follow the discussion in Ref.~[\onlinecite{VolovikBook1}]. Let us imagine that we have two distinct topological insulators in the domains $x>0$ and $x<0$ of the two dimensional plane (assuming the two dimensional cartesian coordinates $x$ and $y$). $x=0$ is the position of the domain wall. Suppose the parameters of the Hamiltonian smoothly change as a function of $x$, so that asymptotically at large positive and large negative $x$ the Hamiltonian looks like that of a topological insulator with the invariant $N_3^R$ and $N_3^L$.  

The Green's function of this topological insulator is no longer just the function of the momenta, as the momentum $k_x$ is no longer a good quantum number. However, the translational invariance in the $y$-direction is still present. Generally, the Green's function, a matrix in the position space, can be written as $G_{ab}(\Omega; x, x'; y-y')$. One can perform a Fourier transform of the Green's function in the $y$-direction, along the domain wall, and a Wigner transform\cite{WignerTransform,Kamenev2004}  of the Green's function in the $x$-direction perpendicular to the domain wall, to arrive at the Green's function $G_{ab}(\Omega, \k, x)$. 
   
With the help of the Wigner transformed Green's function, we define a local $x$-dependent topological invariant ($k_0=\omega$, $k_1=k_x$, and $k_2=k_y$ as before and $\alpha$, $\beta$, $\gamma$ are summed over the values $0$, $1$, and $2$) 
\begin{eqnarray} && N_3(x) =\\ &&     \frac{\epsilon_{\alpha \beta \gamma}}6  \tr \int_{-\infty}^\infty d\omega \int \frac{d^2 k}{(2\pi)^2} \, G^{-1} \d_{k_\alpha} G G^{-1} \d_{k_\beta} G G^{-1} \d_{k_\gamma} G. \nonumber
\end{eqnarray}
Here $G^{-1}(\Omega, \k, x)$ is defined as satisfying the  equation
\be \sum_b G^{-1}_{ab}(\Omega,\k,x) \, G_{bc}(\Omega, \k, x) = \delta_{ac}.
\ee Here, and below, we use the notation tr to denote the trace over matrix indices which do not include coordinates or momenta, and the notation Tr whenever the trace includes the summation over coordinates and momenta, such as in \rfs{eq:tra}. 

Crucially, $G^{-1}$ is no longer the inverse of the Green's function and is distinct from the Wigner transform of the matrix $K$ defined earlier in \rfs{eq:trueinverse}. $K$ at the same time is the true inverse of the Green's function. Far away from the domain wall where the Hamiltonian and the Green's function stop depending on $x$ explicitly, $G^{-1}$ and the Wigner transform of $K$ coincide. 

Since the topological invariant can take integer values only, as $x$ varies we expect that $N_3$ charges abruptly from $N_3^L$ to $N_3^R$ at some value of $x$, say $x=0$. 

Using a technique developed in Ref.~[\onlinecite{VolovikBook1}], we can show that the difference $N_3^R-N_3^L$ is related the topological invariant $N_1$ defined in \rfs{eq:simplestinvariant}. To do that, consider $N_1$ calculated at a certain value of momentum $k_y=\Lambda>0$
\be \label{eq:tra}
  N_1(k_y)  =   {\rm Tr}~\int \frac{d \omega}{\pi i} K(\Omega,k_y)\,  \d_\omega G(\Omega, k_y),
\ee where the symbol Tr denotes summation over all the indices not represented by   $p_y$ and the second line of this equation is added for later convenience. Construct the difference
\be \label{eq:topdif} \oh \left(  \left.  N_1 \right|_{k_y=\Lambda} - \left. N_1 \right|_{k_y=-\Lambda} \right),
\ee where $\Lambda$ is some very large positive momentum. 
For the noninteracting system, this gives the excess of positive energy excitations at $k_y=\Lambda$ over those at $k_y=-\Lambda$, as follows from \rfs{eq:posneg}. If this number is nonzero, this means that there is a branch of excitations which crossed zero as a function of $k_y$. In other words, there are zero energy excitations (edge states) at the boundary between two topological insulators. 

For interacting system, this is a more subtle quantity as we have seen in the previous section. It can be nonzero even if there are no edge states, by moving a zero of the Green's function across zero energy as $k_y$ changes. The formalism we are developing cannot tell which of the two scenarios happen in general, but nevertheless it can relate this number to the difference $N_3^R-N_3^L$. 

To derive this, we need a relation between $K$ and $G^{-1}$. The exact relationship between those is very complicated. If however the $x$-dependence of the Green's function is weak (the Hamiltonian varies slowly as $x$ is taken across the domain wall)  the relationship between them can be derived by the so-called Moyal product expansion or a gradient expansion \cite{Kamenev2004}. This is defined in the following way. Suppose we have two functions $A(x,y)$ and $B(x,y)$, and a third function $C(x,y)$ which is a convolution of the first two,
\be C(x,z) = \int dy \, A(x,y) B(y,z).
\ee
Then the Wigner transform of these functions is approximately related to the Wigner transform of the first two by the following expansion
\begin{eqnarray}  \label{eq:moyal} && C(k,x) = A(k,x)\,  B(k,x) +\\ &&  \frac{1}{2i} \left( \d_x A(k,x) \, \d_k B(k,x) - \d_k A(k,x) \, \d_x B(k,x) \right) + \dots, \nonumber
\end{eqnarray} valid when $x$ dependence of these functions is weak.  

Note that the trace of $C$ can be calculated in a straightforward way
\be \label{eq:traceformula} \int dx \, C(x,x) = \int \frac{dx dk}{2\pi} C(k,x) = \int \frac{dx dk}{2\pi} A(k,x) B(k,x).
\ee

The inverse Green's function $K$'s convolution with the Green's function $G$ is a delta-function. This allows us to calculate $K$ approximately via $G^{-1}$ with the help of the gradient expansion
\begin{eqnarray}  \label{eq:gradexp}   K  &=& G^{-1}+   \frac{1}{2i} \left( G^{-1} \d_x G G^{-1} G_{k_x} G G^{-1} - \right. \cr && \left.  G^{-1} \d_{k_x} G G^{-1} \d_x G G^{-1} \right) + \dots.
\end{eqnarray}

We use this to calculate \rfs{eq:tra} using the gradient expansion. We substitute \rfs{eq:gradexp} into \rfs{eq:tra} (there is no need to use Moyal product expansion in the product of $K$ and $\d_\omega G$, due to \rfs{eq:traceformula}). We find
\begin{eqnarray} && N_1(k_y) =\tr \int \frac{dx dk_x d\omega}{4\pi^2} \left( G^{-1} \d_x G G^{-1} \d_{k_x} G G^{-1} \d_\omega G - \right. \cr
&& \left. G^{-1} \d_{k_x} G G^{-1} \d_{x} G G^{-1} \d_\omega G  \right)= \cr
&&\sum_{\alpha \beta \gamma} \epsilon_{\alpha \beta \gamma} \tr \int \frac{dx dk_x d \omega}{12 \pi^2} G^{-1} \d_\alpha G G^{-1} \d_\beta G G^{-1} \d_\gamma G.
\end{eqnarray}
Here $\alpha$, $\beta$, $\gamma$ go over $x, k_x, \omega$. 
   
The next step relies on the construction of a vector 
\be n_\alpha = \epsilon_{\alpha \beta \gamma \delta} \tr G^{-1}\d_\beta G G^{-1} \d_\gamma G G^{-1} \d_\delta G,
\ee where $\alpha$, $\beta$, $\gamma$, $\delta$ now span the four-dimensional space of $x$, $k_x$, $k_y$, and $\omega$.    This vector satisfies
\be \d_\alpha n_\alpha =0,
\ee as can be checked by direct differentiation. The relevant difference of the invariant \rfs{eq:topdif} can be thought of as the flux of this vector through a three dimensional surface formed by the planes $p_y=\Lambda$ and $p_y=-\Lambda$, surrounding the singular point $\omega=0$, $x=0$, while $k_x$ and $k_y$ are tuned to appropriate values. 

That same flux can be written as a flux through the surfaces $x=L$ and $x=-L$. Finally, this can be brought to the form
\be \label{eq:volrelation} \oh \left( N_1(\Lambda)-N_1(-\Lambda)  \right) = N_3(L)-N_3(-L).
\ee
This  formula relates what happens at the boundary to the difference of topological invariants. In case of noninteracting system, it proves that for two topologically distinct systems bordering each other, there are always boundary zero energy excitations (chiral edge states). In the presence of interactions, it is possible for the left hand side of this formula to be nonzero without having any zero energy excitations at the boundary as we saw before. Whether this happens in practice requires further study of the zeroes of the Green's functions in interacting topological insulators. 

\section{Topological insulators with chiral symmetry}
\label{sec:6}
We now proceed to generalize the previous discussion to the topological insulators with chiral or sublattice symmetry. These include the insulators in classes AIII (sublattice symmetry only), BDI (sublattice symmetry and time reversal invariance), CII (sublattice symmetry and spin orbit coupling), DIII (time reversal invariant superconductors with spin orbit coupling such as phase B of He$^3$) and CI (spin singlet time reversal invariant superconductor), all of which can possess a ${\mathbb Z}$ topological invariant in odd spacial dimensions only, unlike the remaining insulators including those discussed in the previous chapters which could possess a ${\mathbb Z}$ topological invariants only in even spacial dimensions. Expressions for the topological invariants such as \rfs{eq:TKNN} work only in the absence of chiral symmetry (they are designed to work in even number of spacial dimensions). In what follows we propose equivalent expressions for chiral topological insulators and proceed to relate those to zero energy boundary excitations and zeroes of the Green's function.

The simplest of these insulators are the ones in class AIII, since they do not have any symmetries besides the sublattice symmetry. Let us first review the noninteracting topological insulators of this type.

\subsection{Noninteracting chiral topological insulators}

Consider a noninteracting topological insulator in class AIII. A noninteracting fermionic problem is described by its Hamiltonian
\be \label{eq:ham} {\hat H} = \sum_{ij} {\cal H}_{ij} \hat a^\dagger_i \hat a_j,
\ee
where the indices $i$, $j$ refer to points on  a lattice, spin, and/or species of fermions in the problem. Such a system belongs to class AIII if there exist a unitary matrix $\Sigma_{ij}$, \be  \sum_j \Sigma^\dagger_{ij} \Sigma_{jk}= \delta_{ik}, \ee such that
\be \label{eq:sym} \sum_{jk} \Sigma^\dagger_{ij} {\cal H}_{jk} \Sigma_{kl} = - {\cal H}_{il}.
\ee
This Hamiltonian is said to be insulating if it has a single particle gap. 

It is straightforward to see that $\sum_{j} \Sigma_{ij} \Sigma_{jk} = \delta_{ik}$ as a consequence of \rfs{eq:sym}, thus the eigenvalues of $\Sigma$ are either $+1$ or $-1$. 

The existence of this ``chiral" or ``sublattice" symmetry given by \rfs{eq:sym} immediately leads to a number of consequences. For any single particle state $\psi_i$ of energy $\epsilon_n$, satisfying 
\be {\cal H}_{ij} \psi_j =   \epsilon_n \psi_i,\ee  there exists a state $\sum_j \Sigma_{ij} \psi_j$ with energy $-\epsilon_n$. Among the states there may be zero energy single particle states satisfying \be {\cal H}_{ij} \psi_j^{(0)} = 0.\ee Those are also eigenstates of $\Sigma_{ij}$ with the eigenvalues either $+1$ or $-1$. The zero energy eigenstates whose eigenvalue of $\Sigma_{ij}$ is $+1$ are called the right zero modes, and those with the eigenvalue $-1$ the left zero modes,
\be \label{eq:zerononint} \Sigma \psi^{(0)}_L = - \psi^{(0)}_L, \  \Sigma \psi^{(0)}_R =  \psi^{(0)}_R.
\ee

The difference of the number of the right and left zero modes is a topological invariant (this difference cannot change if ${\cal H}_{ij}$ is deformed smoothly). In fact, that difference is called the index of the operator $V_{\alpha \beta}$ from \rfs{eq:c} and the fact that it is a topological invariant is referred to as the ``index theorem". The existence of the index theorem for the chiral topological insulators, as well as its absence in non-chiral topological insulators, in particular in $p+ip$ superconductors, was emphasized in Ref.~\onlinecite{Gurarie2007b}. 

It is convenient to work in the basis of the eigenstates of $\Sigma$. In this basis, $\Sigma$ looks like
\be \label{eq:sigma}
  \Sigma = \left( \matrix { \delta_{\alpha \beta} & 0\cr 0 & -\delta_{\alpha \beta}} \right).
\ee
We often refer to the sites of the lattice which are $+1$ eigenstates of $\Sigma$ as even and $-1$ eigenstates as odd sublattice. 
The Hamiltonian
 takes form
\be \label{eq:c} {\cal H} = \left( \matrix { 0 & V_{\alpha \beta} \cr V^\dagger_{\alpha  \beta} & 0 } \right). 
\ee 
Here $\alpha$ and $\beta$ label sites of each of the sublattices. It is straightforward to check that \rfs{eq:c} is consistent with, and follows from, \rfs{eq:sym}. 

For a translationally invariant system, we can introduce momentum (crystal momentum if on the lattice). In this basis, the Hamiltonian takes the form
\be   \label{eq:ccc} {\cal H} = \left( \matrix { 0 & V_{ab}(\k) \cr V^\dagger_{ab}(\k) & 0 } \right). 
\ee 
Here $a$ and $b$ label remaining states within the lattice basis after the momentum is introduced. 

Such a Hamiltonian is known to be characterized by a topological invariant in any odd dimensional space. The invariant is constructed the following way. $V_{\alpha \beta}$ is a matrix function of momentum $k$. We can construct an expression
\be \label{eq:inv} I_d  \sim  \epsilon_{\alpha_1 \alpha_2 \dots \alpha_d} \tr \int d^d k \prod_{n=1}^d V^{-1} \d_{\alpha_n} V.
\ee Here $\d_{\alpha} = \d_{k_\alpha}$ and appropriate summation is implied. 
The integral goes over all the entire $d$-dimensional space, or, in case if the problem is defined on a lattice, over the Brillouin zone. 
This expression takes on values proportional to an integer, which, mathematically speaking, reflects the fact that $\pi_{d}(GL(M,{\mathbb C}))={\mathbb Z}$  for odd $d$. This invariant is the odd-dimensional counterpart of the TKNN invariant, which exists in the symmetry classes without chiral symmetry only, and is nonzero only in even $d$. 

It is especially simple to see that this is an integer if the matrix $V$ is a one by one matrix, or reduces to a number $z(k)$, and if $d=1$. Then
\be I_1    = \int \frac{dk}{4\pi} \left( \frac 1 z \pbyp{z}{k}  -  \frac 1 {z^*} \pbyp{z^*}{k} \right).
\ee
This measures the number of times $z(k)$ winds around the origin as $k$ goes over the Brillouin zone.

One famous example of the topologically nontrivial insulator in one dimension with the symmetry \rfs{eq:sym} are the Su-Schrieffer-Heeger\cite{Heeger1980} solitons. Indeed, these authors considered a problem of a particle hopping on a 1D lattice with the Hamiltonian
\be \label{eq:ss} \hat H = \sum_{i} \left( t \,  \hat a^\dagger_{2i} \hat a_{2i+1} + t' \, \hat a^\dagger_{2i+1} \hat a_{2i+2} \right) + {\rm h.c.}
\ee
It is straightforward to see that the matrix ${\cal H}$ in this case takes the form
\be {\cal H}(k) = \left( \matrix { 0 & t + t' e^{ik} \cr t+t' e^{-ik} & 0 }\right).
\ee
We see that $t+t' e^{ik}$ winds around the origin if $t'>t$ and does not if $t'<t$. It immediately follows that if for $i>0$, $t>t'$, and for $i<0$, $t<t'$, 
at the boundary between $t>t$ and $t<t'$ insulators there is going to be a zero mode, the Su-Schrieffer-Heeger soliton (this can of course be established also in a direct fashion by constructing the solutions to the Schr\"odinger equation). 

Technically speaking the Hamiltonian \rfs{eq:ss} possesses not only chiral symmetry \rfs{eq:sym} but also time-reversal invariance, so it belongs not to the class AIII, but rather to the class BDI. However, in 1D the distinction between these two classes is not significant (although it becomes  more significant in spaces
of higher dimensionalities).

More generally, any problem where a particle hops on a bipartite lattice possesses the symmetry \rfs{eq:sym} and is said to belong to class AIII (in the absence of any other symmetries), with the matrix $\Sigma$ acting by $+1$ on even and by $-1$ on odd sites, as long as a particle can only hope from an odd site  to an even site or vice versa.

\subsection{Interacting chiral topological insulators}

We would like to see if adding interactions to the Hamiltonian \rf{eq:ham} preserves its topological properties, especially the edge states. To do that, we need to generalize \rfs{eq:sym} to interacting systems. We do it in the way discussed earlier in Sec.~\ref{sec:2}, in particular in Eqs.~\rf{eq:chio}, \rf{eq:chi} and \rf{eq:gsl}.

As a result of the symmetry defined by \rfs{eq:chi} for any state $\left|n\right>$ there exists a state $\hat \Sigma^\dagger \left|n^* \right>$ which has exactly the same energy as the state $\left| n \right>$,
\be \label{eq:symi} \hat H \left| n \right> = E \left| n \right>, \ \hat H \Sigma^\dagger \left| n^* \right> = E \hat \Sigma^\dagger \left| n^* \right>.
\ee
Notice also that if the state $\left| n \right>$ has $N$ particles, then its conjugate state $\hat \Sigma^\dagger \left| n^* \right>$ has $D_f-N$ particles. To see that, we
write
\begin{eqnarray}  \label{eq:partconv} \sum_{i=1}^{D_f} \hat a^\dagger_i \hat a_i \hat \Sigma^\dagger \left| n^* \right> &=& \hat \Sigma^\dagger \hat \Sigma  \sum_{i=1}^{D_f} \hat a^\dagger_i \hat a_i \hat \Sigma^\dagger \left| n^* \right> = \cr
\hat \Sigma^\dagger \sum_{i=1}^{D_f} \hat a_i \hat a_i^\dagger \left| n^* \right> &=& \hat \Sigma^\dagger \left(D_f - \sum_{i=1}^{D_f} \hat a^\dagger_i \hat a_i \right) \left| n^* \right>.
\end{eqnarray}

The lowest energy state of the system obeying this kind of symmetry has exactly $N=D_f/2$ particles, that is, the system is at half filling. Indeed, suppose the number of particles in the lowest energy state is not $D_f/2$. That means that the ground state energy at particle number fixed at $N$, $E_0(N)$, has a minimum at $N \not = D_f/2$. At the same time, it satisfies
\be E_0 \left( N \right)=E_0 \left( D_f-N \right)
\ee
as a consequence of Eqs.~\rf{eq:symi} and \rf{eq:partconv}. 
Therefore, this function also has exactly the same minimum at $N_1=D_f-N$ particles. This implies that the compressibility of the system, $d^2 E_0/dN^2$, is not positive definite. From now one we restrict our attention to systems with positive compressibility, which as we just saw is equivalent to systems at half filling.

An interacting system invariant under \rfs{eq:chi} possesses a topological invariant, a generalization of \rfs{eq:inv} for the case when there are interactions
and reducing to it in case where there are no interactions. To define it, we again introduce the Green's function $G_{ij}(\Omega)$. For a translationally invariant system, it is a function of momentum $\k$. Otherwise, it has the same matrix structure as the noninteracting Hamiltonian \rfs{eq:ccc}.

We construct the following matrix out of the Green's functions (here, as before, $\omega$ is taken to be real)
\be \label{eq:Q} Q(\omega,\k) = G^{-1}(i \omega, \k) \, \Sigma \, G(i \omega,\k).
\ee
Here matrix multiplication is implied,  while $G^{-1}$ is the inverse of the matrix $G(\Omega, \k)$. It is clear that \be Q^2=1.\ee 
We now construct the expression
\be \label{eq:invgr}  I_D \sim  \epsilon_{\alpha_1, \alpha_2, \dots, \alpha_D} \tr \int_0^\infty d\omega \int d^d k \, Q \prod_{n=1}^D \d_{\alpha_n}Q .
\ee
Here $D=d+1$ is the dimension of space-time, $\d_i = \d_{k_i}$ and $\d_0$ implies $\d_\omega$. 
This expression is a topological invariant, which mathematically reflects the fact that $\pi_D\left( U(2M)/U(M) \times U(M) \right) = {\mathbb Z}$ for even $D$. One can check that for the noninteracting systems where $G=\left[ \Omega - \cH \right]^{-1}$, \rfs{eq:invgr} reduces to \rfs{eq:inv}. 

Note that the integration over $\omega$ goes from $0$ to infinity, not from $-\infty$ to $\infty$. The reason for that is that the integral from $-\infty$ to $\infty$ is zero, as can be verified using
\be Q(\omega) = \Sigma^\dagger Q(-\omega) \Sigma,
\ee
which in turn follows from the definition \rfs{eq:Q} as well as the chiral symmetry \rfs{eq:gsl}. 

To understand why \rfs{eq:invgr} is a topological invariant, we need to further elucidate the structure of the chiral Green's function. Taking advantage of the explicit form of the matrix $\Sigma$ given in \rfs{eq:sigma}, we arrive at the following structure
\be \label{eq:gr} G(\Omega) = \left( \matrix { \Delta^{(1)}_{\alpha \beta} (\Omega)  & W_{\alpha \beta} (\Omega) \cr W_{\alpha \beta}^\dagger(\Omega) & \Delta^{(2)}_{\alpha \beta} (\Omega)} \right).
\ee
Here the matrices $\Delta$, $W$ satisfy, thanks to \rfs{eq:gsl}
\be \label{eq:symmetryrelations}  \Delta^{(1,2)}(\Omega) = -\Delta^{(1,2)}(-\Omega), \ W(-\Omega) = W(\Omega).
\ee
For completeness we give the explicit expressions for these matrices in the appendix~\ref{sec:appendixB}. 

For a translationally invariant system, the Green's function can be written in the momentum space as
\be \label{eq:gr25} G(\Omega, \k) = \left( \matrix { \Delta^{(1)}_{ab} (\Omega, \k)  & W_{ab} (\Omega, \k) \cr W_{ab}^\dagger(\Omega, \k) & \Delta^{(2)}_{ab} (\Omega, \k)} \right).
\ee
In the absence of any excitations around zero energy, that is, in case when one deals with an insulator, the functions $\Delta$, $W$, are regular at $\Omega=0$. In particular this means that $\Delta(\Omega=0)$ must vanish, while $W(0)$ is a constant which for now we assume to be nonvanishing. This leads to
\be\label{eq:qasone} Q(\omega \rightarrow 0) \rightarrow  \left( \matrix { -1 & 0 \cr 0 & 1 } \right).
\ee
At the same time, at $\Omega \rightarrow \infty$, the Green's function must behave as $1/\Omega$, so $W$ vanishes in this limit and
\be \label{eq:qastwo} Q(\omega \rightarrow \infty) \rightarrow \left( \matrix { 1 & 0 \cr 0 & -1 } \right).
\ee
Thus the topological invariant measures how $Q$ interpolates between these two values. 

Finally, one can check that \rfs{eq:invgr} is indeed a topological invariant by checking that if $Q$ is varied slightly, to $Q+\delta Q$, \rfs{eq:invgr} does not change. This turns out to be a direct consequence of $Q^2=1$.

The topological invariant \rfs{eq:invgr} is a good starting point to analyze the chiral topological insulators in 3D, including exotic singlet time-reversal invariant superconductors (class CI), Helium III-B (class DIII) and a particle hopping on a bipartite lattice (class AIII) \cite{Schnyder2009,Hosur2010,Volovik2010}. However, in what follows we will use it to analyze chiral problems in one spacial dimension. 


\subsection{Low energy excitations at the boundaries of interacting one dimensional   class AIII topological insulators}

We now have in our possession  the topological invariant for interacting system in chiral symmetry classes. However, what we are really interested in is whether two insulators with two different values of $I$ which are in spacial contact possess  zero energy excitations at the boundary. More precisely, we would like to show that it is possible to add particles without any energy cost at the boundary, just like in our discussion of topological insulators in two spacial dimensions in Sec.~\ref{sec:vol}. Let us show that this is indeed the case. 

From now on we specialize  to one spacial dimension (although generalizations of what follows to higher dimensions should be possible to construct). Suppose we have a 1D interacting chiral topological insulator, whose parameters vary in space (so that it is not translationally invariant). However at far away positive spacial infinity it becomes translationally invariant, with the topological invariant $I_R$, while at far away negative infinity its topological invariant is $I_L$.

Just as we have already discussed in case of two spacial dimensions in Sec.~\ref{sec:vol}, as the system is no longer translationally invariant, its Green's function is no longer a function of momentum only. Nevertheless, we can introduce the Wigner transformed Green's function, the one Fourier transformed with respect to the difference of coordinates. Such a Wigner transformed function depends on both the momentum $k$ and the coordinate $x$. Yet at far away spacial infinity $G$ becomes independent of $x$ and becomes equal to its value for a translationally invariant system. 

We can define $Q$ in the same way as before, 
\be \label{eq:qw} Q(\omega, k, x) = G^{-1}(i\omega, k, x) \ \Sigma \, G(i \omega, k, x).
\ee
Here $G^{-1}(\Omega, k, x)$ is defined as the inverse of the matrix $G(\Omega, k, x)$, which is not the true matrix inverse of $G$. With its help, we can define a position dependent topological invariant $I_2$ by reducing \rfs{eq:invgr} to the case $D=2$ ($d=1$). We denote it by $I$ as opposed to $I_2$ as in this subsection we always work in one dimensional space so using $I$ without specifying dimensionality should not be the cause of confusion.
\be I(x) = \frac{1}{16 \pi i} \tr \int_0^\infty d\omega \int dk \, Q \left( \d_\omega Q \d_k Q - \d_k Q \d_\omega Q \right).
\ee
Then
\be I_L= \left. I \right|_{x=-L}, \ I_R = \left. I \right|_{x=L}
\ee
for some large $L$ such that at $x=L$ or $x=-L$ we are far from the boundary of the two insulators. 

Now we define a vector 
\be n_\alpha = \epsilon_{\alpha \beta \gamma} \tr Q \, \d_\beta Q \d_\gamma Q.
\ee Here $\alpha$, $\beta$, and $\gamma$ go over values $0$, $1$, $2$, which refer to the three effective coordinates, $\omega$, $k$ and $x$.  Notice that
\be \label{eq:diveq} \d_\alpha n_\alpha =0.
\ee 
This important relation follows from the fact  that
\be \label{eq:divergence} \epsilon_{\alpha \beta \gamma} \tr \d_\alpha Q \d_\beta Q \d_\gamma Q =0.
\ee
This in turn can be proving by using that $Q^2 =1$ as well as $\d Q Q=-Q \d Q$. Indeed, 
\begin{eqnarray}
&&  \epsilon_{\alpha \beta \gamma} \tr \d_\alpha Q \d_\beta Q \d_\gamma Q = \cr  &&  \epsilon_{\alpha \beta \gamma} \tr \d_\alpha Q Q^2 \d_\beta Q Q^2 \d_\gamma Q =
 \cr &&
  -\epsilon_{\alpha \beta \gamma} \tr Q \d_\alpha Q  \d_\beta Q Q^2 \d_\gamma Q  Q = \cr &&  - \epsilon_{\alpha \beta \gamma} \tr \d_\alpha Q  \d_\beta Q  \d_\gamma Q. 
\end{eqnarray}
We see that the expression \rfs{eq:divergence} is equal to minus itself, so it is zero.

Similar manipulations can be used to prove that  \rfs{eq:invgr} is indeed a topological invariant, that is, does not change under small deformations in $Q$, as long as $Q^2=1$.

Due to \rfs{eq:diveq}, the following integral is zero
\be \int dS_\alpha n_\alpha=0,
\ee
where the integral is taken over any closed surface in the $\omega$, $k$, $x$ space. In particular, consider a surface formed by the planes $x=L$, $x=-L$  and $\omega=0$ (as well as $\omega=\infty$, where the integral vanishes), with $dS_\alpha$ being the element of this surface. We note that the sum of the integrals over $x=\Lambda$ and $x=-\Lambda$ is equal precisely to $I_R-I_L$. This allows us to express this difference in terms of the integral over the surface $\omega=0$,
or more precisely
\be \label{eq:diff} I_R-I_L =\frac{1}{16 \pi i}\lim_{\omega \rightarrow 0} \tr \int dx dk \, Q  \left( \d_x Q \, \d_k Q - \d_k Q \, \d_x Q \right).
\ee
The limit is needed because at $\omega= 0$ the expression to be integrated is singular as we will see below. 

Now let us relate this difference to the number of zero energy excitations. Earlier when working in 2D we calculated the number of zero energy excitations by a clever application of \rfs{eq:tra}. This is not possible to do in chiral systems however (formally the expression \rfs{eq:tra} is always zero due to \rfs{eq:gsl}). 
Instead we take advantage of a similar, but distinct formula
\be \label{eq:zero} N =-   \lim_{\omega \rightarrow 0} \omega \,  {\rm Tr} \,  \left[ \Sigma K \d_\omega G \right],
\ee
Here $K$ is given by \rfs{eq:trueinverse}. We recall that as before, $K$ coincides with $G^{-1}$ in translationally invariant systems only.

Let us check that \rfs{eq:zero} is indeed equal to the number of zero modes. First of all, in the noninteracting case, $G=\left[i \omega-{\cal H} \right]^{-1}$. This allows us to write
\be \label{eq:zcm} N_0 = \lim_{\omega \rightarrow 0} i \omega \, {\rm Tr} \, \Sigma \left[i \omega-{\cal H} \right]^{-1} = N_R-N_L.
\ee

Here $N_R$ is the number of the right zero modes and $N_L$ is the number of the left zero modes (recall that zero modes satisfy $\Sigma \psi^{(0)}_{R,L}=\pm 
\psi^{(0)}_{R,L}$). This is the correct way of counting zero modes, since it is this difference that is insensitive to small changes in the Hamiltonian as discussed after \rfs{eq:zerononint}.

In case when there are interactions, the analysis of \rfs{eq:zero} is somewhat more involved. Define the eigenvalues and eigenvectors of the Green's function as
in \rfs{eq:eigenvaluegr}, or in other words
\be \sum_j G_{ij}(\Omega)\, \psi_j^{(\mu)}(\Omega) = \lambda_\mu(\Omega)\, \psi_i^{(\mu)}(\Omega).
\ee
The trace in \rfs{eq:zero} can be understood as a sum over these eigenvectors
\be 
 \label{eq:zerotr} N =-   \lim_{\omega \rightarrow 0} \omega \,  \sum_\mu \, \psi^{(\mu)} \,  \Sigma \, K \, \d_\omega G  \, \psi^{(\mu)},
\ee
where matrix multiplication is implied. 

If $\psi(\Omega)$ is an eigenvector of the Green's function $G(\Omega)$ with the eigenvalue $\lambda(\Omega)$, then $\Sigma^\dagger \psi(-\Omega)$ is an eigenvector of the Green's function $G(\Omega)$ with an eigenvalue $-\lambda(-\Omega)$, as a consequence of \rfs{eq:gsl}. 


A special role is played by the eigenvectors $\psi(0)$. Indeed, if it is an eigenvector of $G(0)$, then $\Sigma^\dagger \psi(0)$ is also its eigenvector with the eigenvalue $-\lambda(0)$. These eigenvectors do not contribute to \rfs{eq:zerotr} in the limit of $\omega \rightarrow 0$.

However, there are also the eigenvalues such that $\lambda(\Omega)\sim \Omega$ as $\Omega \rightarrow 0$, or such that $\lambda(\Omega) \sim 1/\Omega$ as $\Omega \rightarrow 0$. Then $\psi(0)$ and $\Sigma^\dagger \psi(0)$ are both eigenfunctions with the same eigenvalue. In that case, just like in the analysis in the previous section \rfs{eq:zerononint}, $\psi(0)$ are the eigenvectors not only of $G$ but also of $\Sigma$, with the eigenvalues $+1$ or $-1$. 

We call the eigenvectors with the eigenvalues $\lambda(\Omega) \sim 1/\Omega$ right and left poles (right poles satisfy $\Sigma \psi_R = \psi_R$, left poles satisfy $\Sigma \psi_L=-\psi_L$), while we call the eigenvectors with the eigenvalues $\lambda(\Omega) \sim \Omega$ right and left zeroes. Then by direct substitution we find, from \rfs{eq:zerotr},
\be N = N_R-N_L - Z_R  + Z_L,
\ee where $Z_R$ and $Z_L$ are the number of right and left zeroes of the Green's function. 


Now we can relate $N$ to the difference between the values of the topological invariant by following the approach described in Sec.~\ref{sec:vol}. This involves expressing the product of two Green's functions  in \rfs{eq:zero} in terms of their Wigner transform using gradient expansion, as well as expressing $K$ in terms of $G^{-1}$ using
\rfs{eq:gradexp}.  This produces two terms. The first one, where we use the gradient expansion while substituting $G^{-1}$ for $K$, looks like
\be-\frac{\omega}{2i} {\rm Tr} \Sigma \left[-G^{-1} \d_x G G^{-1} \d_{\omega, k} G + G^{-1} \d_k G G^{-1} \d_{\omega, x} G \right].
\ee
Note that this term should be zero after integration over $k$ and $x$ due to \rfs{eq:traceformula}. Nevertheless we would like to keep it because it will make the final expression which will be integrated over $k$ and $x$ simpler. 

In the second term we use \rfs{eq:gradexp} for $K$ to find
\begin{eqnarray} &&
- \frac{\omega}{2i} {\rm Tr} \Sigma \left[ G^{-1} \d_x G G^{-1} G_k G^{-1} \d_\omega G  - \right. \cr && \left.  G^{-1} \d_k G G^{-1} G_x G^{-1} \d_\omega G  \right].
\end{eqnarray}
Combining these terms together gives
\begin{eqnarray} \label{eq:nzeromodes} N &=&   \lim_{\omega \rightarrow 0} \omega \, \tr  \Sigma
  \int \frac{dx dk}{4\pi i} 
\left[ G^{-1} \d_x G \, \d_k \left( G^{-1} \d_\omega G \right) -
\right. \cr &&  \left. 
G^{-1} \d_k G\, \d_x \left( G^{-1} \d_\omega G \right) \right].
\end{eqnarray}
At the same time, \rfs{eq:diff} can be rewritten, upon substituting \rfs{eq:qw}, as
\begin{eqnarray} \label{eq:topinvdif} && I_R - I_L = \lim_{\omega \rightarrow 0} \tr \Sigma \int \frac{dx dk}{4\pi i} \, \left(  \d_x G G^{-1} \d_k G G^{-1} -  \right. \cr && \left. \d_k G G^{-1} \d_x G G^{-1}  \right). 
\end{eqnarray}

The expressions Eqs.~\rf{eq:nzeromodes} and \rf{eq:topinvdif} are not identical. Nevertheless, they are related in the following way. The entries of the
Green's function, as introduced in \rfs{eq:gr}, satisfy the relations given by \rfs{eq:symmetryrelations}. As a result, the Green's function has to have one
of the following possible asymptotic behaviors as
$\omega$ is taken to zero. The Green's function might have a pole as $\omega$ is taken to zero at this value of $k$ and $x$ leading to
\be \label{eq:asympole} G(\omega,k,x) \sim \left( \matrix { \frac{\delta^{(1)}_{ab}(k,x)}{i \omega} & W_{ab}(k,x) \cr W^\dagger_{ab}(k,x) & \frac{ \delta^{(2)}_{ab}(k,x)}{i \omega} } \right).
\ee
Alternatively, it is regular
\be \label{eq:asymzero} G(\omega,k,x) \sim \left( \matrix { i \omega \, \delta^{(1)}_{ab}(k,x) & W_{ab}(k,x) \cr W^\dagger_{ab}(k,x) & i \omega \, \delta^{(2)}_{ab}(k,x)  } \right).
\ee
Here $\delta^{(1)}$ and $\delta^{(2)}$ are the appropriate expansion coefficients of $\Delta^{(1)}$ and $\Delta^{(2)}$ from \rfs{eq:gr}.

An interesting feature of both of these expressions is that, in the limit $\omega \rightarrow 0$, $Q$ is either $\Sigma$ or $-\Sigma$, just as in Eqs.~\rf{eq:qasone} and \rf{eq:qastwo} and is $k$ and $x$ independent. As a result, naive substitution of either \rfs{eq:asympole} or \rfs{eq:asymzero} into \rfs{eq:topinvdif} gives zero in the limit of $\omega \rightarrow 0$. However, at the values of $k$ and $x$ where $\delta$ vanish in \rfs{eq:asympole} or $W$ vanish 
in \rfs{eq:asymzero}, $Q$ varies rapidly, giving rise to a delta-function-like contribution to \rfs{eq:topinvdif}. Notice that the values of $k$ and $x$ where $W$ vanishes corresponds to the zero of the Green's function. We mostly elaborate on this argument in two and one dimensional space, although it should be generalizable to higher dimensions.

One can check by a direct substitution that if \rfs{eq:asympole} holds, then \rfs{eq:nzeromodes} is equal to  \rfs{eq:topinvdif}. If, however, \rfs{eq:asymzero} holds
then \rfs{eq:nzeromodes} is equal to minus \rfs{eq:topinvdif}. This was checked for $G$ being a  two by two matrix and  a four by four matrix and we conjecture (although this has not been checked) that this is valid for matrices of all sizes.

As a result, we find that 
\be I_R-I_L = P+Z, \ N = P-Z,
\ee where $P$ is the contribution of the poles and $Z$ is the contribution of the zeroes.  
Since the change in the topological invariant must be an integer, $P$ and $Z$ are also integers. 

Therefore, we conclude that the change in the topological invariant is due to either the poles of the Wigner transformed Green's functions  or the zeroes of the Wigner transformed Green's functions, which occur at some values of $x$ and $k$, where the appropriate values of $x$ are close to the domain wall separating the two topological insulators. The poles' contributions are exactly opposite to those of the zeroes. 

In particular, the topological invariant will change even if there is no poles, $P=0$, as long as there are zeroes present, $Z\not= 0$. Then two topological insulators with distinct topological invariants will have no zero energy excitations at their boundary.    It is tempting to conclude that this scenario is at work for the one dimensional BDI systems discussed in recent papers \cite{Fidkowski2010,Fidkowski2010a,Turner2010}. However, it is not currently clear why the topological invariant considered there had to change by 4 (by 8 if one works with Majorana chains instead of Dirac chains considered here) and not by 1 as would have been natural here (see appendix~\ref{sec:appendixC} to see that $Z$ can be any integer).

\section{Conclusions}

\label{sec:concl}

We have presented a scenario in this paper where an interacting topological insulator can lose its edge states as the interactions are turned on. The method developed here has not yet been developed to a point where it allows to understand when this actually happens. Instead, it explains the mechanism by which this can happen. One should also be careful in interpreting the results of this formalism.  The edge states are understood here as poles in the single particle Green's functions. If there is transport along the edge in such a way that the elementary particle carrying the current along the edge are collective excitations, this formalism may interpret it as the absence of the edge states. 

In this work we restricted our attention to the topological insulators with the topological invariant $\mathbb Z$. The relationship between the invariant and the edge states was explored in two and one spacial dimensions only. It is straightforward to generalize this to higher number of dimensions, in particular to three dimensional space. It is also important to generalize this formalism to the invariant $\mathbb Z_2$, especially in view of current interest to 2D and 3D AII topological insulators. The expression for the $\mathbb Z_2$ topological invariant in terms of the Green's functions was found in Ref.~[\onlinecite{Zhang2010}]. What remains is to find what it tells us about the edge states (and zeroes of the Green's functions). All of this will be the subject of future work.

The author is grateful to A. Altland and A. W. W.  Ludwig for many stimulating discussions in the course of this work, and would like to thank S. Ryu, C. L. Kane, L. Radzihovsky, M. Hermele and E. Altman for useful comments. Some of this work was done during visits to the Institute for Theoretical Physics at the University of Cologne, as well as to the Kavli Institute for Theoretical Physics in Santa Barbara. This work was supported by the NSF grants DMR-0449521, PHY-0904017, and PHY-0551164.

\appendix
\section{Alternative generalization of chiral symmetry in the presence of interactions}
\label{sec:appendixA}
Instead of \rfs{eq:chi}, we could use the following construction to generalize the chiral symmetry to the interacting case. Consider an operator $\hat \Sigma$ such that
\be \label{eq:altchi} \hat \Sigma^\dagger \hat H \hat \Sigma = - \hat H.
\ee
This operator does not mix creation and annihilation operators, so that its action on those operators is simply
\be \hat \Sigma^\dagger \hat a_i \hat \Sigma  = \Sigma_{ij} \hat a_j,  \ \hat \Sigma^\dagger \hat a_i^\dagger \hat \Sigma  =  \hat a_j^\dagger \Sigma^\dagger_{ji}.
\ee
We can check that noninteracting chiral Hamiltonians automatically satisfy \rfs{eq:altchi} due to \rfs{eq:sym}. An example of an interacting Hamiltonian in 1D invariant under this operation would be (here we have spinful fermions hopping on a 1D lattice)
\begin{eqnarray}  \hat H &=& \sum_{i, \sigma=\uparrow, \downarrow} t_i \left( \hat a^\dagger_{\sigma i} \hat a_{\sigma, i+1} + \hat a^\dagger_{\sigma, i+1} \hat a_{\sigma i} \right) + \\
&& U \sum_i \left( \hat a^\dagger_{\uparrow i+1} \hat a_{\downarrow i}^\dagger \hat a_{\downarrow i} \hat a_{\uparrow i}
+ \hat a^\dagger_{\uparrow i} \hat a^\dagger_{\downarrow i} \hat a_{\downarrow i} \hat a_{\uparrow i+1} \right). \nonumber
\end{eqnarray}
Indeed, it changes sign under the sign change of every creation and annihilation operator on odd sites of the lattice. 

As a result of the symmetry \rfs{eq:altchi}, for every state $\left| n \right>$ there exist a state $\hat \Sigma \left| n \right>$ whose energy is exactly opposite to that of $\left| n \right>$. In particular, in addition to the ground state $\left| 0 \right>$, the system should have a maximally excited state $\left| \tilde 0 \right>$, with the energy opposite to that of the ground state. 

We can then define two Green's functions. One, denoted by $G(\Omega)$ is defined with respect to the ground state. The other, denoted $\tilde G(\Omega)$, is defined with respect to the maximally excited state. One can now check that the sum of these two functions, \be \label{eq:a4} G^s(\Omega) = \oh \left( G(\Omega) + \tilde G(\Omega) \right) \ee satisfies the same relationship \rfs{eq:gsl} which Green's functions for the chiral systems are supposed to satisfy. It follows that this is a legitimate generalization of the chiral symmetry to the interacting case, distinct from \rfs{eq:chi}. It is possible to prove that in the absence of interactions \rfs{eq:a4} coincides with the usual Green's function, but this does not have to be true in the presence of interactions. We leave the question whether any interesting models of this type can be constructed and studied  to further work. 

\section{Spectral decomposition of the chiral Green's functions}
\label{sec:appendixB}

The general spectral decomposition \rfs{eq:grleh} valid for any Green's function can be further specialized to the case where chiral symmetry is present. In the presence of chiral symmetry, any state $\left| n \right>$ with one extra particle compared to the ground state acquires a partner, a state $\hat \Sigma^\dagger \left| n^* \right>$ with one less particle than the ground state. As a result, the spectral decomposition becomes
\be G_{ij}(\Omega) = \sum_n \frac{\left< 0 \right| \hat a_i \left| n \right> \left< n \right| \hat a^\dagger_j \left| 0 \right>}{\Omega- \epsilon_n} +
  \frac{\left< 0 \right| \hat a^\dagger_j \hat \Sigma^\dagger \left| n^* \right> \left< n^* \right| \hat \Sigma \, \hat a_i \left| 0 \right>}{\Omega+ \epsilon_n}.
\ee
Here $\left| n \right>$ are the states with one extra particle compared to the ground state, while $\epsilon_n = \omega_n^+>0$, defined in \rfs{eq:omegaplus}. 
The second term in this expression can be transformed in the following way
\begin{eqnarray}
\left< 0 \right| \hat a^\dagger_j \hat \Sigma^\dagger \left| n^* \right> \left< n^* \right| \hat \Sigma \, \hat a_i \left| 0 \right> &=& \cr
\left< 0 \right| \hat \Sigma^\dagger \hat \Sigma \hat a^\dagger_j \hat \Sigma^\dagger \left| n^* \right> \left< n^* \right| \hat \Sigma \hat a_i  \hat \Sigma^\dagger \hat \Sigma \left| 0 \right>& =& \cr
\left< n \right| \hat \Sigma^\dagger \, \hat a_j \hat \Sigma \hat \Sigma^\dagger \left| 0^*\right> \left< 0^* \right| \hat \Sigma \hat \Sigma^\dagger \hat a_i^\dagger \hat
\Sigma \left| n \right>&  = & \cr
\Sigma_{il} \left<0 \right| \hat a_l \left| n \right> \left< n \right| \hat a^\dagger_k \left| 0 \right> \Sigma^\dagger_{kj}. &&
\end{eqnarray}
To go from the second to the third line, \rfs{eq:transp} was used, while in the final line takes advantage of Eqs.~\rf{eq:chio} and \rf{eq:grchi}. 
Finally, this gives
\begin{eqnarray} \label{eq:decomp} && G_{ij}(\Omega) =\\ &&  \sum_n \left[ \frac{\left< 0 \right| \hat a_i \left| n \right> \left< n \right| \hat a^\dagger_j \left| 0 \right>}{\Omega- \epsilon_n} +
 \frac{\Sigma_{il} \left< 0 \right| \hat a_l \left| n \right> \left< n \right| \hat a^\dagger_k \left| 0 \right> \Sigma^\dagger_{kj} }{\Omega+ \epsilon_n} \right]. \nonumber
\end{eqnarray}
Note that this is compatible with the constraint \rfs{eq:gsl}. 

With the help of this expression, we can write the following explicit construction for matrices defined in \rfs{eq:gr}.
We introduce the matrix element
\be U_{in} = \left< 0 \right| \hat a_i \left| n \right>.
\ee
We also employ the notations $U^{(1)}_{\alpha n}$ and $U^{(2)}_{\alpha n}$, which are equal to $U_{in}$ as $i$ belongs to the first or the second sublattice. 
Now \rfs{eq:decomp} gives
\begin{eqnarray}  \label{eq:explicit} \Delta^{(1)}_{\alpha \beta} &=&  \sum_n U^{(1)}_{\alpha n} {U^{(1)}}^\dagger_{n \beta}\frac{2\Omega} {\Omega^2- \epsilon_n^2}, \cr
 \Delta^{(2)}_{\alpha \beta} &=&  \sum_n U^{(2)}_{\alpha n} {U^{(2)}}^\dagger_{n \beta}\frac{2\Omega} {\Omega^2- \epsilon_n^2},\cr
 W_{\alpha \beta} &=&  \sum_n U^{(1)}_{\alpha n} {U^{(2)}}^\dagger_{n\beta}\frac{2\epsilon_n} {\Omega^2- \epsilon_n^2}.
\end{eqnarray}

\section{Zeroes of the chiral Green's functions and the topological invariant}
\label{sec:appendixC}

Suppose the chiral Green's function takes the form
\be \label{eq:gsim2} G(\omega) = \left( \matrix { i \omega  \delta^{(1)} & W \cr W^* & i \omega  \delta^{(2)} } \right),
\ee
where $\delta^{(1)}$ and $\delta^{(2)}$ are real positive functions of $x$ and $k$ while $W$ is a complex function
of $x$ and $k$. This is a 2 by 2 version of the Green's function \rfs{eq:asymzero}. Let us calculate the
change in the topological invariant due to this form of the Green's function. Substituting this into \rfs{eq:topinvdif} we find
\be \label{eq:ontheway}
I_R-I_L = \lim_{\omega \rightarrow 0} \int \frac{dk dx}{2\pi i}\frac{ \omega^2 \delta^{(1)} \delta^{(2)} \left( \d_p W \d_x W^* - \d_x W \d_p W^* \right)}{ \left( \delta^{(1)} \delta^{(2)} \omega^2 + W W^* \right)^2}.
\ee
Only the terms with nonzero $\omega \rightarrow 0$ limit are shown. 

Suppose $W$ vanishes at some special values of $k$ and $x$. At this point, in the limit $\omega \rightarrow 0$, the Green's function obviously vanishes, or as we say, has a zero.

We write $W=X+i Y$ where $X$ and $Y$ are real and they both vanish at those same values of $k$ and $x$. Then
\be
I_R-I_L = \lim_{\omega \rightarrow 0} \int \frac{dk dx}{\pi }\frac{ \omega^2 \delta^{(1)} \delta^{(2)} \left(\d_x X \d_k Y - \d_k X \d_x Y \right)}{ \left( \delta^{(1)} \delta^{(2)} \omega^2 +X^2+Y^2 \right)^2}.
\ee
As can be checked in a straightforward manner,
\be \lim_{\omega \rightarrow 0} \frac{\omega^2}{\left(\omega^2 + X^2 +Y^2\right)^2} = \pi \, \delta(X) \delta(Y). 
\ee 
This gives
\be I_R-I_L = \pm \int dX dY \delta(X) \delta(Y) = \pm 1.
\ee
One can check that the result of substitution of the form \rfs{eq:gsim2} into \rfs{eq:nzeromodes} gives exactly the same expression as  \rfs{eq:ontheway} except with the opposite sign.

\bibliography{draft}

\begin{thebibliography}{27}
\bibitem[{Wig()}]{WignerTransform}
\bibinfo{note}{A Wigner transform of a function of two variables $f(x,x')$ is
  defined as $f(k, R) = \int dr \, e^{i k r} f(R+r/2, R-r/2)$.}
\end{thebibliography}

\end{document}